\documentclass[showpacs,aps,twocolumn]{revtex4}
\usepackage{epsfig}
\usepackage{graphicx}
\usepackage{amsmath,amssymb,amsfonts}
\usepackage{array}
\usepackage{url}
\usepackage{hyperref}
\usepackage{multirow}
\usepackage{float}
\usepackage{lineno}
\usepackage{xspace}
\usepackage[usenames,dvipsnames]{color}

\newcommand{\bef}{\begin{figure}}
\newcommand{\eef}{\end{figure}}
\newcommand{\bc}{\begin{center}}
\newcommand{\ec}{\end{center}}

\newcommand{\be}{\begin{equation}}
\newcommand{\ee}{\end{equation}}
\newcommand{\bea}{\begin{eqnarray}}
\newcommand{\eea}{\end{eqnarray}}

\def\ba{\begin{eqnarray}}
\def\ea{\end{eqnarray}}
\begin{document}
\title{Possible Formation of a Perfect Fluid in $pp$, $p$-Pb, Xe-Xe and Pb-Pb Collisions at the Large Hadron Collider Energies: A Color String Percolation Approach}

\author{Dushmanta Sahu}
\author{Sushanta Tripathy}
\author{Raghunath Sahoo\footnote{Corresponding Author Email: Raghunath.Sahoo@cern.ch, Presently CERN Scientific Associate at CERN, Geneva, Switzerland}}
\affiliation{Department of Physics, Indian Institute of Technology Indore, Simrol, Indore 453552, India}
\author{Swatantra Kumar Tiwari}
\affiliation{Department of Applied Science and Humanities, Muzaffarpur Institute of Technology, Muzaffarpur- 842003, India}

\begin{abstract}
Isothermal compressibility ($\kappa_{\rm T}$) is an important thermodynamic observable which gives information about
the deviation of a system from perfect fluid behavior. In this work, for the first time we have estimated the isothermal compressibility of QCD matter formed in high energy hadronic and nuclear collisions using color string percolation model (CSPM), where we investigate the change in $\kappa_{\rm T}$ as a function of final state charged particle multiplicity and initial percolation temperature across various collision species. The estimated initial percolation temperature for different collision systems at different collision energies helps us to have a better understanding of the system at the initial phase of evolution. The comparison of the CSPM results for isothermal compressibility with that for the well known fluids, indicates that the matter formed in heavy-ion collisions might be the {\it closest perfect fluid} found in nature. This estimation complements the well-known observation of minimum shear viscosity to entropy density ratio for a possible QGP medium created in heavy-ion collision experiments. A threshold of pseudorapidity density of charged particles, $\langle dN_{\rm ch}/d\eta \rangle \geq 20 $ in the final state event multiplicity is observed, after which one may look for a possible QGP formation at the LHC energies.

\end{abstract}
\date{\today}
\maketitle

\section{Introduction}
\label{intro}
One of the most stimulating areas of ongoing research is about the lesser known state of matter called quark gluon plasma (QGP). It is a deconfined state of matter having partons (quarks and gluons) as the degrees of freedom. The Large Hadron Collider (LHC) at CERN, Switzerland, is at the forefront of the endeavour to expand our knowledge horizon in this exciting direction of physics. QGP is governed by Quantum Chromodynamics (QCD) and is the result of a first order/crossover phase transition from normal baryonic matter. The ADS/CFT calculation provides a lower bound (KSS bound) to the shear viscosity to entropy density ratio for any fluid found in nature. The elliptic flow measurements from heavy-ion collisions at Relativistic Heavy Ion Collider (RHIC) at BNL found that the medium formed in such collisions are closer to the KSS bound, which might indicate QGP as nearly a perfect fluid~\cite{Kovtun:2004de,Biro:2011bq,Plumari:2011re,starNPA}. Studying the various thermodynamic properties of QGP is thus very crucial to have a proper understanding of this new state of matter. 

Isothermal compressibility ($\kappa_{\rm T}$) is another important thermodynamic observable which gives information about the deviation of a real fluid from a perfect fluid. A perfect fluid should be incompressible, i.e. $\kappa_{\rm T}$ = 0. $\kappa_{\rm T}$ describes the relative change in the volume of a system with the change in the pressure at constant temperature. To have the proper idea about the equation of state (EOS) of the system, $\kappa_{\rm T}$ has utmost importance. $\kappa_{\rm T}$ is linked to number density fluctuations and it can be expressed in terms of free energy with respect to pressure. Back in 1998, it was proposed that the estimation of isothermal compressibility of matter produced in high energy collisions
would be possible on an event-by-event basis with a simultaneous measurement of particle multiplicity, volume and temperature of the system \cite{Mrowczynski:1997kz}.  For a second order phase transition, $\kappa_{\rm T}$ is expected to show a divergence~\cite{Mukherjee:2017elm}. It would be interesting to see how the values of $\kappa_{\rm T}$ for a QGP medium compares with well-known fluids we deal with in day-to-day life. Recent measurements at the LHC suggest possible formation of QGP-droplets~\cite{ALICE:2017jyt,Khachatryan:2016txc,Bjorken:2013boa,Sahoo:2019ifs} in smaller systems such as $pp$ and $p$-Pb collisions. The calculation of $\kappa_{\rm T}$ for smaller collision systems would have a significant contribution to the above argument. The only way to estimate such a fundamental thermodynamic quantity, $\kappa_{\rm T}$ using experimental inputs is through a theoretical model, and here we have used the established CSPM model \cite{Braun:2015eoa}, which has been very successful
in describing a deconfinement phase transition at the extreme conditions of temperature and energy density. In a recent work~\cite{Mukherjee:2017elm}, $\kappa_{\rm T}$ is studied as a function of collision energies for the hadronic degrees of freedom using available experimental data, various transport models and hadron resonance gas (HRG) model. The estimations of transport models and HRG model show a decrease of $\kappa_{\rm T}$ as a function of collision energy, particularly at lower center-of-mass energies. We have also studied the effect of Hagedorn mass spectrum on $\kappa_{\rm T}$ using HRG model~\cite{Khuntia:2018non}, which lowers down the values of $\kappa_{\rm T}$ particularly at higher collision energies. Recently, there are experimental explorations on the possible direct estimation of isothermal compressibility by the ALICE experiment at the LHC \cite{ALICE:2021hkc}. However, this measurement gives a conservative
upper limit of $\kappa_{\rm T}$ for central Pb-Pb collisions at $\sqrt{s_{\rm NN}}$ = 2.76 TeV because of various reasons like
additional uncorrelated particle production and non-thermal sources. 

At high-energies, multi-particle production can be explained in terms of color string percolation model (CSPM). In this model, color strings are stretched between the target and the projectile particles and the final hadrons are produced by the hadronization of these strings \cite{Braun:2015eoa}. Color flux tubes are assumed to be stretched between the colliding partons in terms of gluon fields, which are restricted to some transverse space radius. The strings have some finite area in the transverse space. The number of strings grow as the energy of the collision and the number of colliding partons increase, and they start overlapping in the transverse space. After a certain critical string density, ($\xi_{\rm c}$), a macroscopic cluster appears which marks the percolation phase transition \cite{Braun:2015eoa}. In some of the recent works it has been suggested that in heavy-ion collisions, fast thermalization can occur through the existence of an event horizon which is caused by the rapid deceleration of the colliding nuclei~\cite{Castorina:2007eb,Bylinkin:2014vra}. In such cases, the thermalization can happen through Hawking-Unruh effect~\cite{hawk,unru}. As suggested by Hawking, the black holes can evaporate by quantum pair production and behave as if they have an effective temperature, $T_{H} = \frac{1}{8\pi GM}$, where $1/(4GM)$ is the acceleration due to gravity at the surface of a black hole of mass $M$. Unruh showed that a similar type of effect can arise in a uniformly accelerated frame, in which an observer outside the system can detect a thermal radiation with a temperature, $T = a/2$, $a$ being the acceleration. Similarly in CSPM, the strong color field inside the large cluster causes deceleration of the $q\bar q$ pair which can be perceived as a thermal temperature because of Hawking-Unruh effect \cite{hawk,unru}. 
In this work, we have estimated the isothermal compressibility of matter formed in high energy collisions using CSPM for the first time. Here, we have taken advantage of the usefulness of color string percolation and found out how $\kappa_{\rm T}$ changes as a function of the final state charged particle multiplicity across various collision species at the LHC energies. We have also estimated the initial percolation temperature for different collision systems at different collision energies.

In view of particle production being driven by the final state multiplicity across the collision species starting from pp, p-Pb, Xe-Xe and
 Pb-Pb collisions at the LHC energies, it is worth going for a comprehensive study of system thermodynamics using experimental inputs as a function of final
 state event multiplicity. However, as it is known due to the low density of particles in the phase space and the physical interpretation of the
 corresponding observables, should be done with caution.
 
The paper is organized as follows. Sec.~\ref{formulation} briefly describes the basic ingredients of CSPM and encompasses the formulation of isothermal compressibility in the CSPM. In Sec.~\ref{res}, the results obtained using the formulation are discussed and finally Sec.~\ref{sum} summarizes the findings of the work.

\section{Formulation}
\label{formulation}
To formulate the isothermal compressibility in color string percolation approach, we first discuss the basic ingredients of CSPM formalism \cite{Braun:1999hv}. 

\begin{figure}[ht!]
\begin{center}
\includegraphics[scale = 0.5]{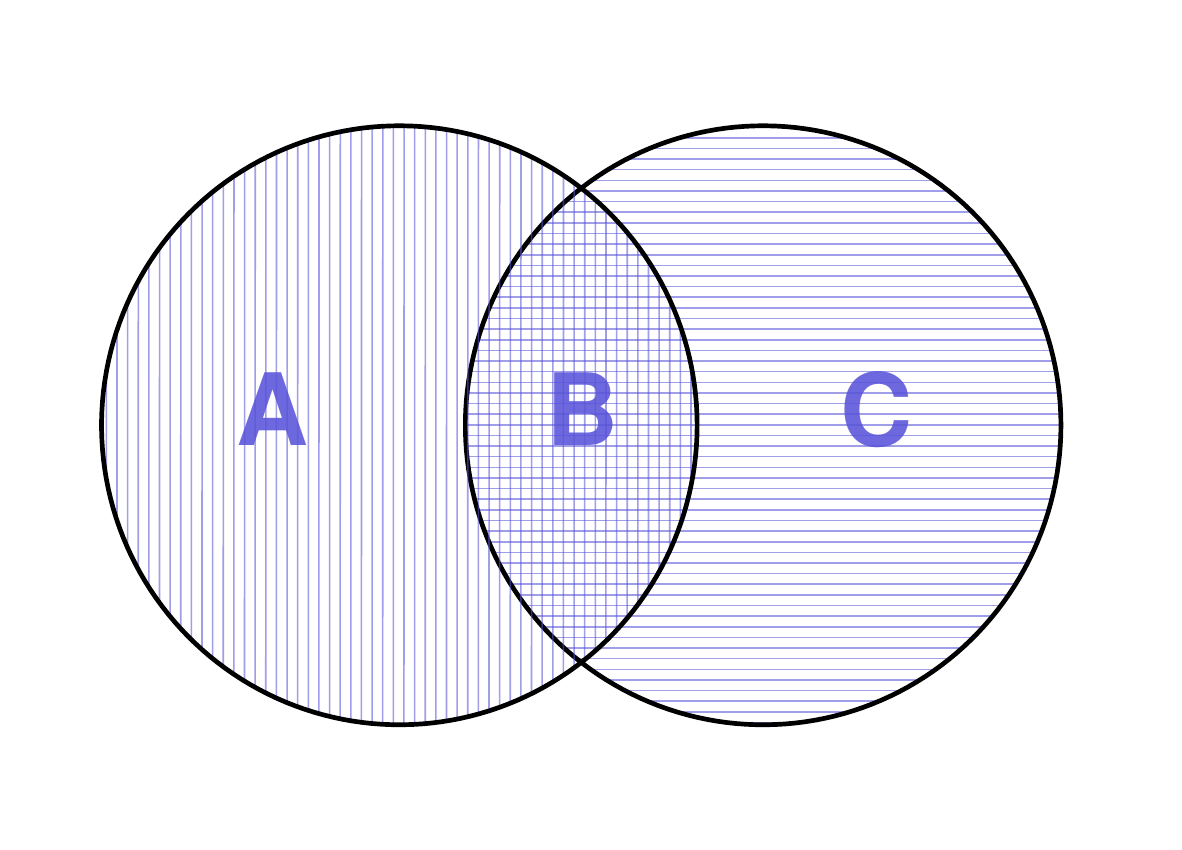}
\caption{(Color Online) Projection of two overlapping strings on the transverse plane.}
\label{fig1}
\end{center}
\end{figure}


Let us assume two color strings, each with a transverse area of $S_1$. They overlap with each other partially in the area $S_B$ and their non-overlapping area is $S_A (\equiv S_C)$. The string multiplicity $\mu_n$ and the average transverse momentum squared $\langle p_{T}^{2} \rangle$ are related to the string's field strength and consequently to the string's generating color. With the assumption of independent emission from the three regions A, B and C in Fig. 1, one can write \cite{Braun:1999hv},

\begin{equation}
\mu_n/\mu_1=2(S_A/S_1)+\sqrt{2}\,(S_B/S_1),
\end{equation}
where $\mu_1$ is a multiplicity for a single string. 
To obtain $\langle p_{T}^{2} \rangle$, the total transverse momentum squared of
all observed particles is divided by the total multiplicity. Thus, for a cluster of two
strings one can get \cite{Braun:1999hv}, 
\begin{equation}
\frac{\langle p_{T}^{2} \rangle}{\langle p_{T}^{2}\rangle_{1}}=\frac{2(S_A/S_1)+\sqrt{2}\sqrt{2}\,(S_B/S_1)}{2(S_A/S_1)+\sqrt{2}\,(S_B/S_1)} \nonumber
\end{equation}
\begin{equation}
=\frac{2}{2(S_A/S_1)+\sqrt{2}(S_B/S_1)},
\end{equation}
where, $\langle p_{T}^{2}\rangle_{1}$ is the average transverse momentum squared for a single string. 

After generalizing to $N_{s}$ number of overlapping strings, the total 
multiplicity can be written as \cite{Braun:1999hv},
\begin{equation}
\mu_n/\mu_1=\sum_i\sqrt{n_i}\,(S^{(i)}/S_1),
\end{equation}
where, the sum is over all individual overlaps $i$ of $n_i$ strings with  
areas $S^{(i)}$.
Similarly, for the $\langle p_{T}^{2} \rangle$ we can write,
\begin{equation}
\frac{\langle p_{T}^{2}\rangle}{\langle p_{T}^{2} \rangle_{1}}=\frac{\sum_i n_i\,(S^{(i)}/S_1)}{\sum_i\sqrt{n_i}\,
(S^{(i)}/S_1)}=
\frac{N_{s}}{\sum_i\sqrt{n_i}\,(S^{(i)}/S_1)}.
\end{equation}

In order to calculate the sums over $i$, one needs to identify all individual overlaps of the strings 
with their own areas. We combine all the terms with $n_i=n$ overlapping strings
into a single term, which sums all such overlaps into a total area of 
$n$ overlapping strings $S_{\rm N}$. Now,
\begin{equation}
\label{5}
\mu_n/\mu_1=\sum_{n=1}^{N_{s}}\sqrt{n}\,(S_{\rm N}/S_1)
\end{equation}
and
\begin{equation}
\langle p_{T}^{2}\rangle/\langle p_{T}^{2}\rangle_{1}=
\frac{N_{s}}{\sum_{n=1}^{N_{s}}\sqrt{n}\,(S_{\rm N}/S_1)}.
\end{equation}

Now, let's assume the projections of the strings onto the
transverse space are distributed uniformly in the interaction
area $S_n$ with a density $\rho$. Introducing a dimensionless parameter \cite{Braun:1999hv},
\begin{equation} \xi=\rho S_1 = \frac{N_{s} S_1}{S_{\rm N}}.
\end{equation}


 For Pb-Pb, Xe-Xe and $p$-Pb collisions, the values of $S_{\rm N}$ are taken from Ref.~\cite{Loizides:2017ack}. While for the $pp$ collisions, the radius is calculated from IP-Glasma model and then is used to estimate the interaction area~\cite{McLerran:2013oju}. 
 
The thermodynamic limit suggests the number of the strings
$N_{s} \rightarrow\infty $ while $\xi$ is  fixed. In this limit, the distribution of the overlaps of $n$ strings is Poissonian having a 
mean value $\xi$, 
\begin{equation}
\label{10}
p_n= \frac{\xi^n}{n!}e^{-\xi}.
\end{equation}
From Eq.\ref{5}, we observe that the multiplicity is damped as a result of overlapping by a factor
\begin{equation}
F(\xi)=\frac{\mu_n}{N_{s}\mu_1}=\frac{\langle\sqrt{n}\rangle}{\xi},
\end{equation}
where, the average is taken over the Poissonian distribution.

From Eq.\ref{10}, the fraction of the total area occupied by the strings is given by
\begin{equation}
\label{eq11}
\sum_{n=1}p_n=1-e^{-\xi}.
\end{equation}

The above equation after being divided by $\xi$ gives the compression factor. According to CSPM picture, the
multiplicity is damped by the square root of the compression factor. Thus, the damping factor or the color suppression factor is given by \cite{Braun:1999hv}
\begin{equation}
F(\xi)=\sqrt{\frac{1-e^{-\xi}}{\xi}}.
\end{equation}

To evaluate the initial value of $\xi$, we fit the experimental data of $pp$ collisions at $\sqrt{s} = 200$ GeV by using the following function~\cite{Srivastava:2011vz}:
\begin{equation}
\label{eq2}
\frac{d^{2}N_{\rm ch}}{dp^{2}_{T}} = \frac{a}{(p_{0}+p_{\rm T})^\alpha},
\end{equation}
where $a$ is the normalization factor and $p_{0}$ and $\alpha$ are the fitting parameters given as, $p_{0} = 1.982$ and $\alpha = 12.877$~\cite{Braun:2015eoa}. Here, $p_{0}$ serves as a low momentum cut-off which is decided from fitting power law to the momentum spectra. In order to evaluate the interaction of strings in $pp$, $p$-A and A+A collisions at the LHC energies, we update the parameter $p_{0}$ as \cite{Braun:2015eoa},
\begin{equation}
\label{eq3}
p_{0} \rightarrow p_{0} \bigg(\frac{\langle \frac{N_{\rm s}S_{1}}{S_{\rm N}}\rangle}{\langle \frac{N_{\rm s}S_{1}}{S_{\rm N}}\rangle_{\rm pp}}\bigg)^{1/4}.
\end{equation}

In the thermodynamic limit i.e. $N_{\rm s}$ and $S_{\rm N}$ $\rightarrow$ $\infty$ and keeping $\xi$ fixed, we get \cite{Scharenberg:2018oyj}
\begin{equation}
\label{eq4}
\bigg\langle\frac{N_{\rm s}S_{1}}{S_{\rm N}}\bigg\rangle = \frac{1}{F^{2}(\xi)},
\end{equation}

Using Eq.\ref{eq2}, for $pp$, $p$-Pb, Xe-Xe and Pb-Pb collision systems at the LHC energies we obtain,
\begin{equation}
\label{eq6}
\frac{d^{2}N_{ch}}{dp^{2}_{T}} = \frac{a}{(p_{0}\sqrt{F(\xi)_{pp, \sqrt{s}=200 ~\rm GeV}/F(\xi)_{pp,pA,AA}}+p_{T})^{\alpha}}.
\end{equation}
In low energy $pp$ collisions, we assume $\big\langle \frac{N_{\rm s}S_{1}}{S_{\rm N}}\big\rangle \sim 1$, due to low string overlap probability \cite{Tarnowsky:2007nj}. We have used Eq. \ref{eq6} to fit the soft part of the $p_{\rm T}$ spectra with the $p_{\rm T}$ range 0.12-1.0 GeV/c.
The initial temperature of the percolation cluster can be defined in terms of $F(\xi)$ as \cite{Hirsch:2018pqm,Sahoo:2017umy,Mishra:2020epq},
\begin{equation}
\label{eq7}
T(\xi) = \sqrt{\frac{\langle p^{2}_{T}\rangle_{1}}{2F(\xi)}}.
\end{equation}
By using $T_{c} = 167.7\pm2.8$ MeV \cite{Becattini:2010sk,Braun:2015eoa} and $\xi_{\rm c} \sim 1.2$ \cite{Isichenko:1992zz,Braun:2015eoa}, we get $\sqrt{\langle p^{2}_{\rm T}\rangle_{1}} = 207.2\pm3.3$~MeV, from which we can get the single string-squared average momentum, $\langle p^{2}_{\rm T}\rangle_{1}$. By using this value in Eq. \ref{eq7}, we can get the initial temperature for different $F({\xi})$ values.

To calculate the isothermal compressibility of different systems, we proceed as follows. From basic thermodynamics, the isothermal compressibility is given as,
\begin{equation}
\label{eq8}
\kappa_{\rm T} = -\frac{1}{V}\frac{\partial V}{\partial P}\bigg\vert_{T},
\end{equation}
where, $V, ~P$ and $T$ are volume, pressure and temperature of the system, respectively. 

To express this thermodynamic quantity in terms of the CSPM parameter we write, 
\begin{equation}
\label{eq9}
\kappa_{\rm T} = -\frac{1}{V}\frac{\partial V}{\partial \xi}\frac{\partial \xi}{\partial P}
\Rightarrow \kappa_{\rm T} = -\frac{1}{V}\frac{\partial V}{\partial \xi} \frac{1}{\frac{\partial P}{\partial \xi}}.
\end{equation}
The volume in this case can be defined as $V = S_{\rm N}L~$\cite{Sahoo:2019xjq}, where $L$ is the longitudinal dimension of the string~($\sim 1$~fm)~\cite{Wong}. Also, pressure can be defined as, $P = (\epsilon - \Delta T^{4})/3$, where $\epsilon$ is the energy density given by Bjorken hydrodynamics and $\Delta$ is the trace anomaly \cite{Sahoo:2017umy}. The energy density is given by \cite{Sahoo:2018dcz}, 
\begin{equation}
\label{eq10}
\epsilon = \frac{3}{2}\frac{\frac{dN_{\rm ch}}{dy}\langle m_{\rm T}\rangle}{S_{\rm N} \tau_{\rm pro}}
\end{equation}
where $m_T$ = $\sqrt{m^2+p_T^2}$ is the transverse mass with m being the mass of pion as pion is the most abundant in a multiparticle production process, and $\tau_{\rm pro}$ is the parton production time, which is assumed as~$\sim$ $\frac{2.405\hbar}{\langle m_{\rm T}\rangle}$~\cite{Wong}.
Pressure can be simplified as,
\begin{equation}
\label{eq11}
P = \frac{\langle m_{\rm T}\rangle dN_{ch}/dy}{2\tau_{\rm pro}S_{\rm N}} - \frac{\Delta T^{4}}{3}.
\end{equation}
In CSPM, the trace anomaly can be expressed as the inverse of shear viscosity to entropy density ratio~\cite{Scharenberg:2018oyj};
\begin{equation}
\label{eq12}
\Delta \simeq \frac{1}{\eta/s} = \frac{5(1-e^{-\xi})}{TL}.
\end{equation}
Thus, the expression of pressure changes to,
\begin{equation}
\label{eq13}
P = \frac{\langle m_{\rm T}\rangle dN_{ch}/dy}{2\tau_{\rm pro}S_{\rm N}} - \frac{5T^{3}(1-e^{-\xi})}{3L}.
\end{equation}
By using these expressions in Eq.\ref{eq9} and simplifying we finally get,
\begin{equation}
\label{eq14}
\kappa_{\rm T} = \frac{1}{\frac{\langle m_{\rm T}\rangle dN_{ch}/dy}{2\tau_{\rm pro}S_{\rm N}}- \frac{5T^{3}e^{-\xi}\xi}{3L}}. 
\end{equation}
Here, $dN_{\rm ch}/dy$ is considered to be $dN_{\rm ch}/d\eta$ assuming a Bjorken correlation of rapidity and pseudo-rapidity. The values of $\langle dN_{\rm ch}/d\eta \rangle$, which are the event classifiers in hadronic and nuclear collisions, are taken from Refs.~\cite{Acharya:2019bli,Acharya:2018orn,Abelev:2013vea,Acharya:2019yoi,Acharya:2019rys,Acharya:2018hhy,Acharya:2020zji}. By using Eq.~\ref{eq14}, we have computed the isothermal compressibility for different collision systems at different collision energies. Keeping in mind that the study of multiplicity dependence of isothermal compressibility is our prime focus while looking for a threshold in final state charged particle multiplicity for the formation of QGP, let us now discuss the results in the next section.

\section{Results and Discussion}
\label{res}

\begin{figure}[ht!]
\begin{center}
\includegraphics[scale = 0.45]{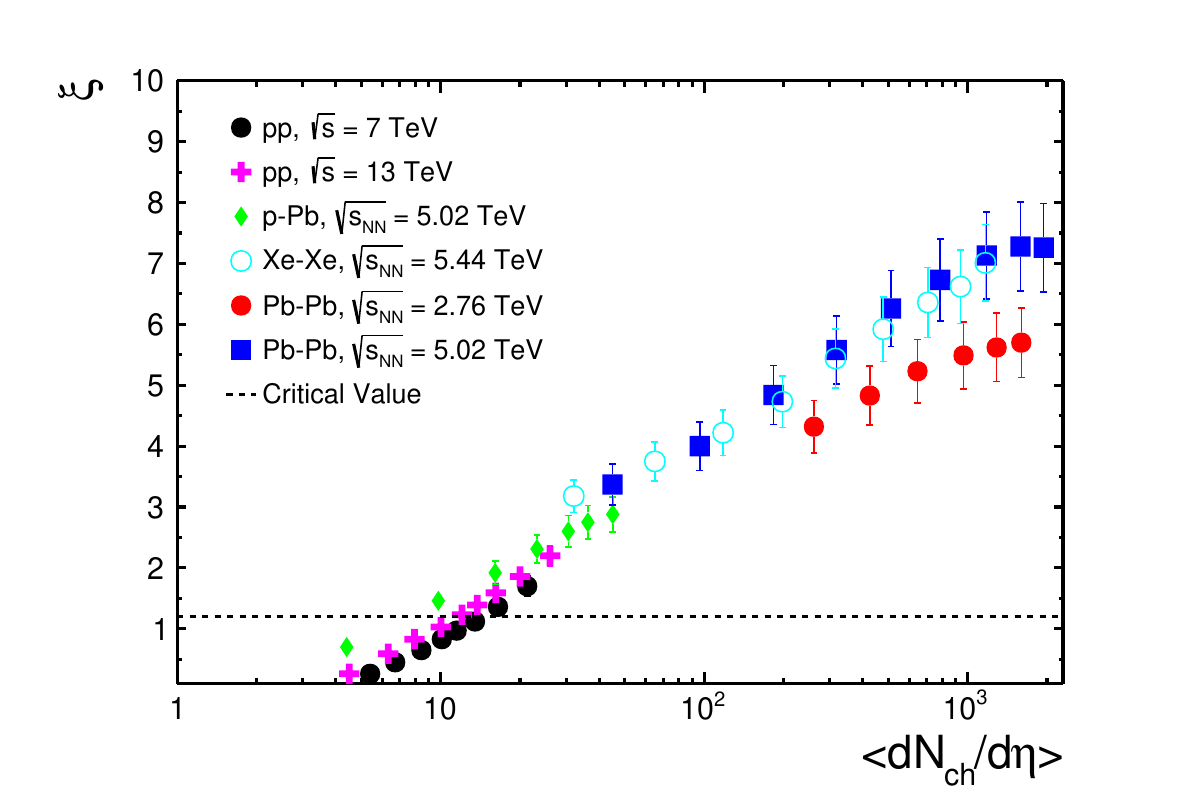}
\caption{(Color Online) Percolation parameter ($\xi$) as a function of charged particle multiplicity for $pp$ collisions at $\sqrt{s}$ = 7 and 13 TeV, $p$-Pb collisions at $\sqrt{s_{\rm NN}}$ = 5.02 TeV,  Xe-Xe collisions at $\sqrt{s_{\rm NN}}$ = 5.44 TeV and Pb-Pb collisions at $\sqrt{s_{\rm NN}}$ = 2.76 and 5.02 TeV. The dotted line represents the critical string density after which a macroscopic cluster appears.}
\label{fig1}
\end{center}
\end{figure}

\begin{figure}[ht!]
\begin{center}
\includegraphics[scale = 0.45]{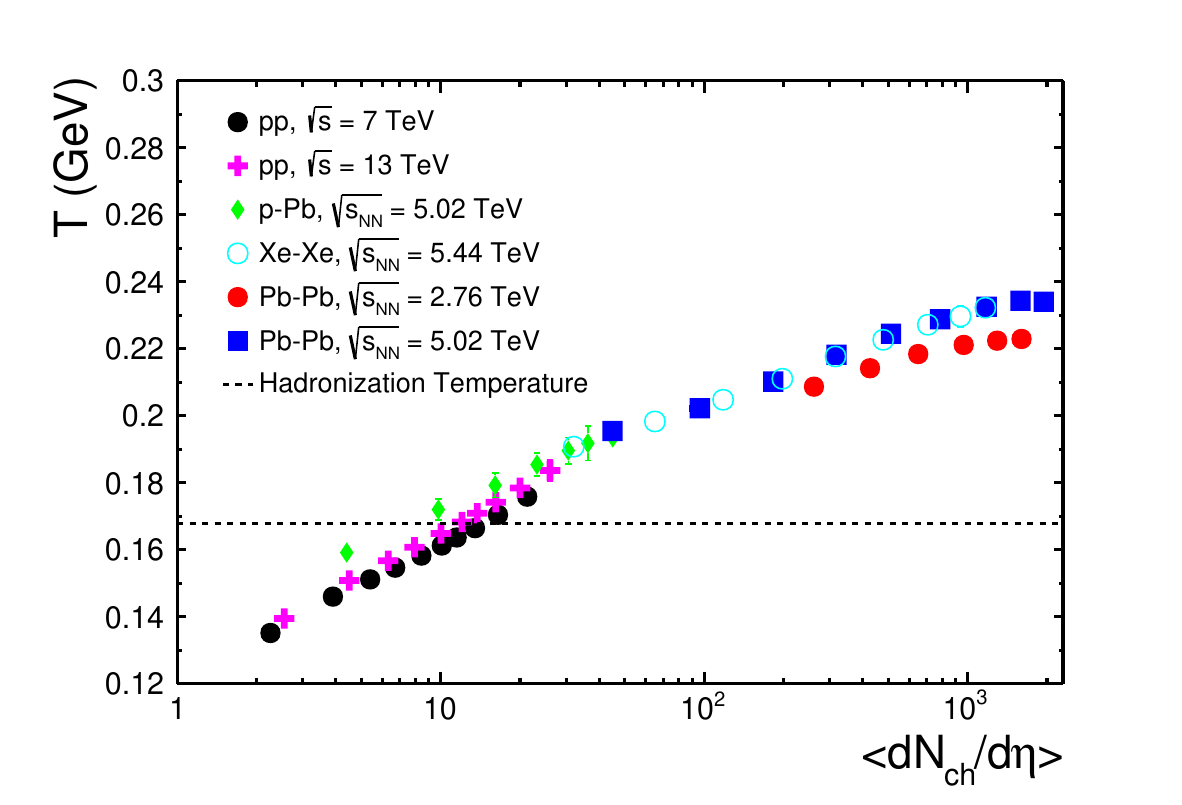}
\caption{(Color online) Initial temperature as a function of charged particle multiplicity for for $pp$ collisions at $\sqrt{s}$ = 7 and 13 TeV, $p$-Pb collisions at $\sqrt{s_{\rm NN}}$ = 5.02 TeV, Xe-Xe collisions at $\sqrt{s_{\rm NN}}$ = 5.44 TeV and Pb-Pb collisions at $\sqrt{s_{\rm NN}}$ = 2.76 and 5.02 TeV. The dotted line represents the reported hadronization temperature~\cite{Becattini:2010sk}.}
\label{fig2}
\end{center}
\end{figure}

\begin{figure*}[ht!]
\centering
\includegraphics[scale = 0.44]{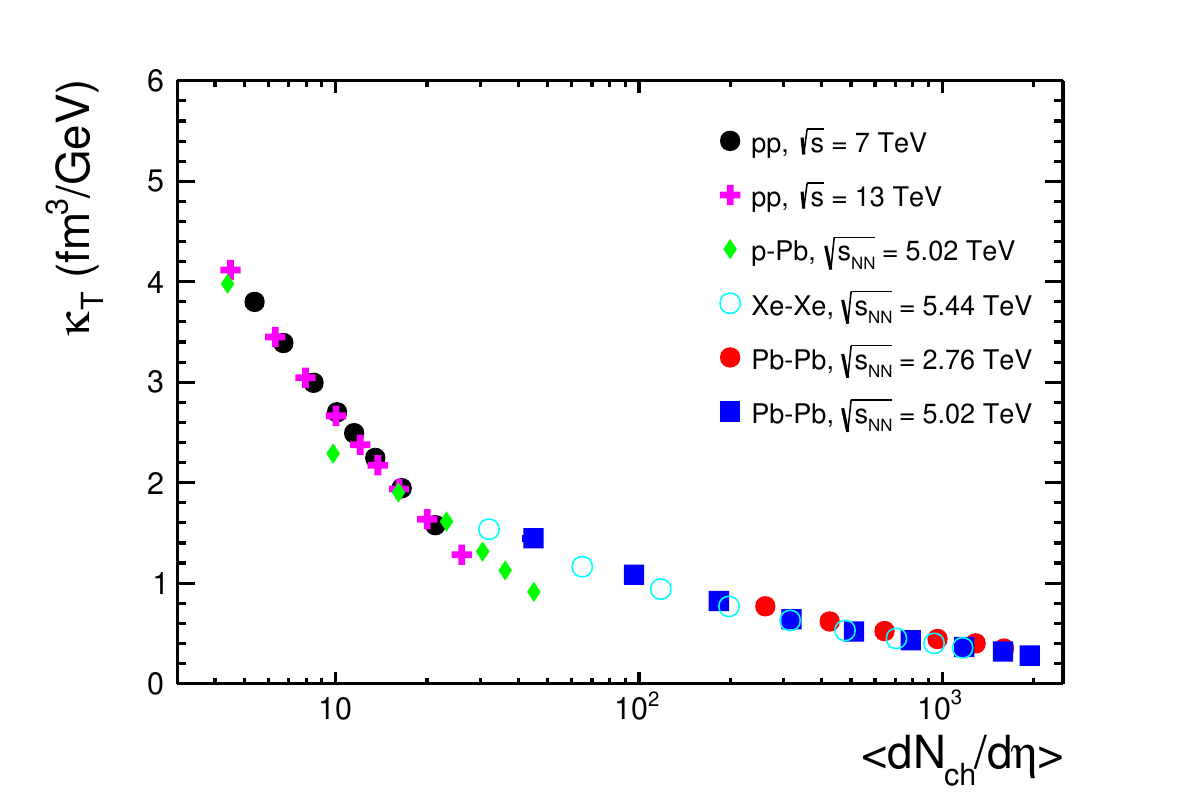}
\includegraphics[scale = 0.44]{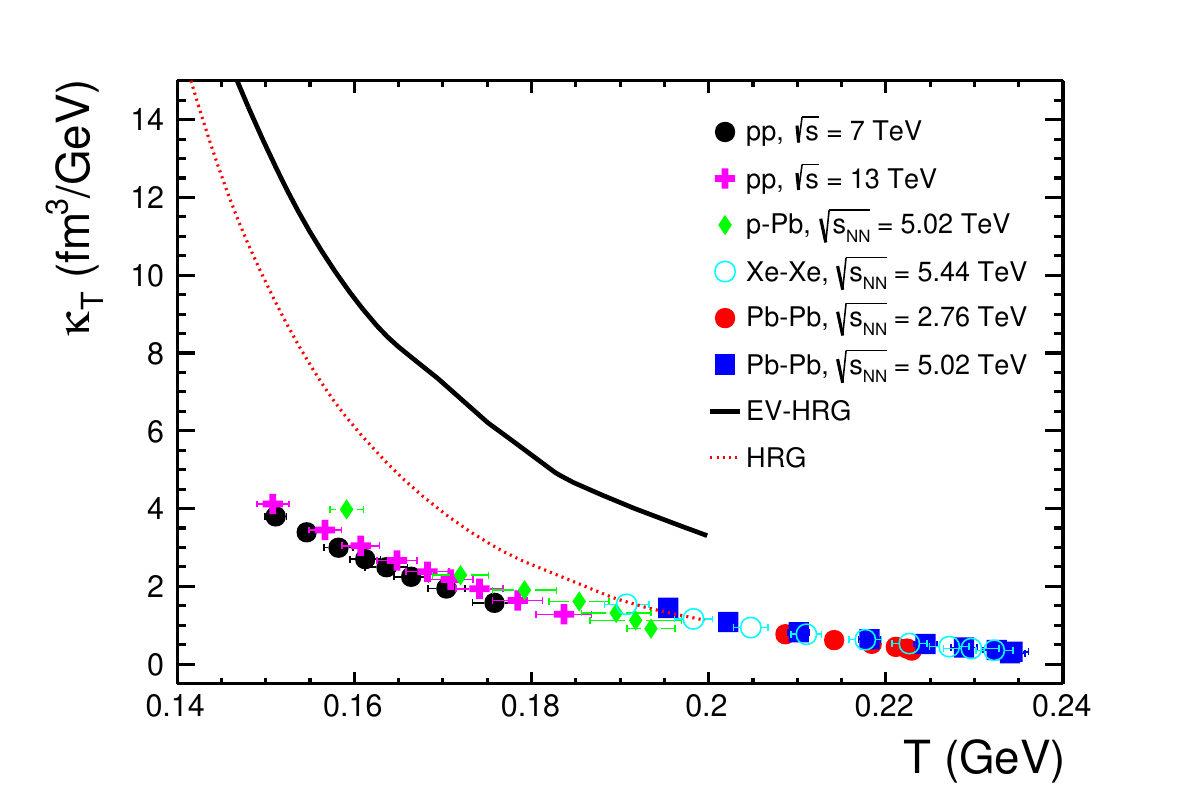}
\caption{(Color Online) $\kappa_{\rm T}$ as a function of charged particle multiplicity (left panel) and temperature (right panel) for $pp$ collisions at $\sqrt{s}$ = 7 and 13 TeV, $p$-Pb collisions at $\sqrt{s_{\rm NN}}$ = 5.02 TeV, Xe-Xe collisions at $\sqrt{s_{\rm NN}}$ = 5.44 TeV and Pb-Pb collisions at $\sqrt{s_{\rm NN}}$ = 2.76 and 5.02 TeV. The dotted red line represents the values from HRG model and the solid black line represents the values from EV-HRG model \cite{Khuntia:2018non}.}
\label{fig3}
\end{figure*}

Figure~\ref{fig1} shows the variation of percolation density parameter ($\xi$) as a function of charged particle multiplicity ($\langle dN_{\rm ch}/d\eta \rangle$). We clearly see $\xi$ is strongly dependent on $\langle dN_{\rm ch}/d\eta \rangle$ and increases smoothly with the increase in the charged particle multiplicity. This suggests that the string percolation density is responsible for the production of final state particles. More is the density, the higher is the number of hadrons produced. We also see that for different collision systems such as $pp$ and Pb-Pb, the evolution is not continuous, although the evolution trend is the same. This indicates that the collision system and collision species have roles to play for particle production along with final state charged-particle multiplicity.  $\xi_{\rm c} = 1.2$ denotes the critical value of this parameter, at which the first order phase transition is assumed to occur~\cite{Braun:2015eoa}. We observe that, beyond $\langle dN_{\rm ch}/d\eta \rangle \simeq$ 20, the value of $\xi$ is found to be more than the critical value, which indicates that a percolation phase transition might have occurred for these events. This can be linked to a possible QGP formation and the threshold in pseudorapidity density of charged particles is seen to be, $\langle dN_{\rm ch}/d\eta \rangle \geq$ 20, which is consistent with the recent findings~\cite{Sharma:2018jqf,Sahu:2019tch,Sahoo:2019ifs}. In addition, we have reported the color suppression factor ($F(\xi)$) in Ref.~\cite{Sahu:2020mzo} for all the collision systems and collision energies mentioned here.

In Fig.~\ref{fig2}, we have plotted the initial percolation temperature ($T$) as a function of final state charged particle multiplicity. As seen in the previous figure, we see a clear dependency of~$T$ on $\langle dN_{\rm ch}/d\eta \rangle$. For large values of charged particle multiplicity, we observe higher initial percolation temperature. This is a confirmation of the fact that the initial energy density must be high for large charged particle multiplicity, thus the initial percolation temperature will be high 
as well. This is also true in Bjorken's hydrodynamic picture and also from experimental measurements of initial energy density \cite{Bjorken:1982qr,Adler:2004zn,Adams:2004cb,Adam:2016thv}. The dotted line represents the hadronization temperature obtained from statistical thermal model~\cite{Becattini:2010sk}, which is also around the critical temperature predicted by lattice QCD calculations for a deconfinement transition \cite{Borsanyi1,Borsanyi2}. The initial temperature for the events with $\langle dN_{\rm ch}/d\eta \rangle$ $\leq$ 20 has lower values compared to the hadronization temperature, which could be an indication of no-QGP scenario.

By using the Eq.~\ref{eq12}, we have estimated the isothermal compressibility for different collision systems at different collision energies. We have plotted $\kappa_{\rm T}$ as a function of charged particle multiplicity in the left panel of Fig.~\ref{fig3}. It is observed that $\kappa_{\rm T}$ decreases rapidly for low $\langle dN_{\rm ch}/d\eta \rangle$ and becomes very small and remains nearly constant for higher $\langle dN_{\rm ch}/d\eta \rangle$. The charged particle multiplicity (density in the phase space), $\langle dN_{\rm ch}/d\eta \rangle \geq 20$ is the threshold after which the isothermal compressibility becomes the lowest and almost stays flat as a minima. This threshold corresponds to the initial
temperature, which is higher than the critical temperature, $T_c$. This observation could be compared with the $T/T_c$ = 1 in Fig. 2 of Ref. \cite{Stokic:2008jh}, where the inverse dimensionless compressibility, $R_k$ (scaled with Gibbs free energy density) should vanish at $\mu/T \simeq$  1, $\mu$ being the quark chemical potential. The value of $R_k$ at $T/T_c$ shows a minima, which also goes in line with our observation of a minima in isothermal compressibility for the threshold charged particle density of $\langle dN_{\rm ch}/d\eta \rangle \geq 20$. However, it should also be mentioned here that our estimations are for zero baryon density at the LHC energies.
The observed behavior of $\kappa_{\rm T}$ is very similar to the theoretical predictions for a hot and dense hadronic matter obtained at the RHIC and LHC~\cite{Kakati:2017xvr}. Isothermal compressibility is plotted as a function of temperature in the right panel of Fig.~\ref{fig3}. We have also added the values of isothermal compressibility from HRG and EV-HRG model\cite{Khuntia:2018non}  to have a better comparison. We see that, for higher temperature (after the hadronization temperature), the value of $\kappa_{\rm T}$ becomes the minimum, regardless of the collision systems.

For liquids, the value of isothermal compressibility is expected to be small because a unitary change in pressure causes a very small change in volume. On the contrary, for a gaseous system, the isothermal compressibility is expected to be higher. Figure~\ref{fig3} suggests that at highest charged particle multiplicity, the value of isothermal compressibility is lowest and close to zero. This could be related to liquid-like behavior for a QCD matter. However, the events with low charged-particle multiplicity show rapid change in the values of $\kappa_{\rm T}$ compared to the ones at higher charged-particle multiplicity. This could be an indication of possible change of dynamics in the system. Similar values of $\kappa_{\rm T}$ for high multiplicity $pp$, $p$-Pb, Xe-Xe and Pb-Pb collisions show an indication of similar dynamics and possibly similar system formation. The reported values of isothermal compressibility for water at room temperature and mercury are $\sim$ $6.62\times10^{42}~\rm{fm^3/GeV}$~\cite{water} and $\sim$ $5.33 \times 10^{41}~\rm{fm^3/GeV}$~\cite{mercury}, respectively, which are very high compared to the value obtained for the QCD matter for the most central Pb-Pb collisions. This is an indication of QGP being a closest perfect fluid found in nature. This measurement nicely complements the measurements of lowest shear viscosity to entropy density ratio for a possible QGP medium \cite{Kovtun:2004de,Biro:2011bq,Plumari:2011re,starNPA}. In addition, this is also supported by similar 
estimations using CSPM approach \cite{Hirsch:2018pqm}.\\


\section{Summary}
\label{sum}
In summary,

\begin{enumerate}
\item From the study of isothermal compressibility of QCD matter, the value of $\kappa_{\rm T}$ for Quark Gluon Plasma is found to be very less compared with water and mercury. This could be an indication of QGP being the closest perfect fluid found in nature.

\item The current measurement of isothermal compressibility complements the measurements of lowest shear viscosity to entropy ratio for a possible QGP medium.


\item Similar values of $\kappa_{\rm T}$ for high multiplicity $pp$, $p$-Pb, Xe-Xe and Pb-Pb collisions show an indication of similar dynamics and possibly similar medium formation. 

\item A threshold in pseudorapidity density of charged particles is found to be $\langle dN_{\rm ch}/d\eta \rangle \geq$ 20 for a possible QGP formation at the LHC energies.
\end{enumerate}
The observations made in this work are of paramount importance in view of the search for QGP droplets in high-multiplicity $pp$ collisions at the LHC energies.



\vspace{10.005em}

  
\begin{thebibliography}{}

\bibitem{Kovtun:2004de} 
  P.~Kovtun, D.~T.~Son and A.~O.~Starinets,
  Phys.\ Rev.\ Lett.\  {\bf 94}, 111601 (2005).
  
  \bibitem{Biro:2011bq} 
  T.~S.~Biro and E.~Molnar,
  Phys.\ Rev.\ C {\bf 85}, 024905 (2012).
 
  \bibitem{Plumari:2011re} 
  S.~Plumari and V.~Greco,
  AIP Conf.\ Proc.\  {\bf 1422}, 56 (2012).

 \bibitem{starNPA} J. Adam {\it et al.} [STAR Collaboration], Nucl.\ Phys.\ A {\bf 757}, 102 (2005).
 
\bibitem{Mrowczynski:1997kz}
S.~Mr\'{o}wczy\'{n}ski,
Phys. Lett. B \textbf{430}, 9 (1998).

\bibitem{Mukherjee:2017elm}
  M.~Mukherjee, S.~Basu, A.~Chatterjee, S.~Chatterjee, S.~P.~Adhya, S.~Thakur and T.~K.~Nayak,
  Phys.\ Lett.\ B {\bf 784}, 1 (2018).

\bibitem{ALICE:2017jyt} 
  J.~Adam {\it et al.} [ALICE Collaboration],
  Nature Phys.\  {\bf 13}, 535 (2017).
  
  \bibitem{Khachatryan:2016txc} 
  V.~Khachatryan {\it et al.} [CMS Collaboration],
  Phys.\ Lett.\ B {\bf 765}, 193 (2017).
  
  \bibitem{Bjorken:2013boa} 
  J.~D.~Bjorken, S.~J.~Brodsky and A.~Scharff Goldhaber,
  Phys.\ Lett.\ B {\bf 726}, 344 (2013).
  
  \bibitem{Sahoo:2019ifs}
  R.~Sahoo,
  AAPPS Bull.\  {\bf 29}, 16 (2019).
  
\bibitem{Braun:2015eoa} 
  M.~A.~Braun, J.~Dias de Deus, A.~S.~Hirsch, C.~Pajares, R.~P.~Scharenberg and B.~K.~Srivastava,
  Phys.\ Rept.\  {\bf 599}, 1 (2015).
  
  \bibitem{Khuntia:2018non} 
  A.~Khuntia, S.~K.~Tiwari, P.~Sharma, R.~Sahoo and T.~K.~Nayak,
  Phys.\ Rev.\ C {\bf 100}, 014910 (2019).
  
\bibitem{ALICE:2021hkc}
S.~Acharya \textit{et al.} [ALICE Collaboration],
Eur. Phys. J. C \textbf{81}, 1012 (2021).
  
\bibitem{Castorina:2007eb}
P.~Castorina, D.~Kharzeev and H.~Satz,
Eur. Phys. J. C \textbf{52}, 187 (2007).

\bibitem{Bylinkin:2014vra}
A.~A.~Bylinkin, D.~E.~Kharzeev and A.~A.~Rostovtsev,
Int. J. Mod. Phys. E \textbf{23}, 1450083 (2014).

\bibitem{hawk} 
S. W. Hawking,  Commun. Math. Phys. Appl. Math. {\bf 43}, 199 (1975).

\bibitem{unru} 
W. G. Unruh,  Phys. Rev. D {\bf 14}, 870 (1976).

\bibitem{Braun:1999hv}
M.~A.~Braun and C.~Pajares,
Eur. Phys. J. C \textbf{16}, 349 (2000).
  
  \bibitem{Loizides:2017ack} 
  C.~Loizides, J.~Kamin and D.~d'Enterria,
  Phys.\ Rev.\ C {\bf 97}, 054910 (2018)
  Erratum: [Phys.\ Rev.\ C {\bf 99}, 019901 (2019)].
  
    
  \bibitem{McLerran:2013oju}
L.~McLerran, M.~Praszalowicz and B.~Schenke,
Nucl. Phys. A \textbf{916}, 210 (2013).

\bibitem{Srivastava:2011vz} 
  B.~K.~Srivastava,
  Nucl.\ Phys.\ A {\bf 862-863}, 132 (2011).
  
    \bibitem{Scharenberg:2018oyj}
R.~P.~Scharenberg, B.~K.~Srivastava and C.~Pajares,
Phys. Rev. D \textbf{100}, 114040 (2019).
  
  
  
\bibitem{Tarnowsky:2007nj} 
  T.~Tarnowsky, R.~Scharenberg and B.~Srivastava,
  Int.\ J.\ Mod.\ Phys.\ E {\bf 16}, 1859 (2007).
  
  
    \bibitem{Hirsch:2018pqm} 
  A.~S.~Hirsch, C.~Pajares, R.~P.~Scharenberg and B.~K.~Srivastava,
  Phys.\ Rev.\ D {\bf 100}, 114040 (2019).
  
  
\bibitem{Sahoo:2017umy} 
  P.~Sahoo, S.~K.~Tiwari, S.~De, R.~Sahoo, R.~P.~Scharenberg and B.~K.~Srivastava,
  Mod.\ Phys.\ Lett.\ A {\bf 34}, 1950034 (2019).
  
  \bibitem{Mishra:2020epq}
A.~N.~Mishra, G.~Pai\'c, C.~Pajares, R.~P.~Scharenberg and B.~K.~Srivastava,
Eur. Phys. J. A \textbf{57}, 245 (2021).

  \bibitem{Becattini:2010sk} 
  F.~Becattini, P.~Castorina, A.~Milov and H.~Satz,
  Eur.\ Phys.\ J.\ C {\bf 66}, 377 (2010).
  
  \bibitem{Isichenko:1992zz}
M.~B.~Isichenko,
Rev. Mod. Phys. \textbf{64}, 961 (1992).
  
\bibitem{Sahoo:2019xjq} 
  P.~Sahoo, R.~Sahoo and S.~K.~Tiwari,
  Phys.\ Rev.\ D {\bf 100}, 051503 (2019).
  
    
\bibitem{Wong} C. Y. Wong, Introduction to High Energy Heavy Ion Collisions (World Scientific, Signapore, 1994).
  
\bibitem{Sahoo:2018dcz} 
  P.~Sahoo, S.~De, S.~K.~Tiwari and R.~Sahoo,
  Eur.\ Phys.\ J.\ A {\bf 54}, 136 (2018).

  
\bibitem{Acharya:2019bli}
S.~Acharya \textit{et al.} [ALICE Collaboration],
Phys. Lett. B \textbf{807}, 135501 (2020).
  
  \bibitem{Acharya:2018orn} 
  S.~Acharya {\it et al.} [ALICE Collaboration],
  Phys.\ Rev.\ C {\bf 99}, 024906 (2019).
  
  \bibitem{Abelev:2013vea} 
  B.~Abelev {\it et al.} [ALICE Collaboration],
  Phys.\ Rev.\ C {\bf 88}, 044910 (2013).
  
\bibitem{Acharya:2019yoi}
S.~Acharya \textit{et al.} [ALICE],
Phys. Rev. C \textbf{101}, no.4, 044907 (2020).
  
  \bibitem{Acharya:2019rys} 
  S.~Acharya {\it et al.} [ALICE Collaboration],
  Phys.\ Lett.\ B {\bf 800}, 135043 (2019).
    
    \bibitem{Acharya:2018hhy} S.~Acharya {\it et al.} [ALICE Collaboration], 
  Phys.\ Lett.\ B {\bf 790}, 35 (2019).

\bibitem{Acharya:2020zji}
S.~Acharya \textit{et al.} [ALICE Collaboration],
Eur. Phys. J. C \textbf{80}, 693 (2020).


    
  \bibitem{Sharma:2018jqf} 
  N.~Sharma, J.~Cleymans, B.~Hippolyte and M.~Paradza,
  Phys.\ Rev.\ C {\bf 99}, 044914 (2019).
 
  
  \bibitem{Sahu:2019tch} 
  D.~Sahu, S.~Tripathy, G.~S.~Pradhan and R.~Sahoo,
 Phys.\ Rev.\ C {\bf 101}, 014902 (2020).
 
 \bibitem{Sahu:2020mzo}
D.~Sahu and R.~Sahoo,
J. Phys. G \textbf{48}, 125104 (2021).
  
    \bibitem{Bjorken:1982qr} 
  J.~D.~Bjorken,
  Phys.\ Rev.\ D {\bf 27}, 140 (1983).
  
  \bibitem{Adler:2004zn} 
  S.~S.~Adler {\it et al.} [PHENIX Collaboration],
  Phys.\ Rev.\ C {\bf 71}, 034908 (2005), 
  Erratum: [Phys.\ Rev.\ C {\bf 71}, 049901 (2005)].
  \bibitem{Adams:2004cb} 
  J.~Adams {\it et al.} [STAR Collaboration],
  Phys.\ Rev.\ C {\bf 70}, 054907 (2004).
  
  \bibitem{Adam:2016thv} 
  J.~Adam {\it et al.} [ALICE Collaboration],
  Phys.\ Rev.\ C {\bf 94}, 034903 (2016).
  

  
   \bibitem{Borsanyi1} S. Borsanyi {\it et al.} J.\ High\ Energ.\ Phys.\ {\bf 2010}, 77 (2010).
\bibitem{Borsanyi2} S. Borsanyi {\it et al.} Phys.\ Lett.\ B {\bf 730}, 99 (2014).
  

\bibitem{Stokic:2008jh} 
  B.~Stokic, B.~Friman and K.~Redlich,
  Phys.\ Lett.\ B {\bf 673}, 192 (2009).
  
  \bibitem{Kakati:2017xvr} 
  S.~K.~Tiwari, S.~Tripathy, R.~Sahoo and N.~Kakati,
  Eur.\ Phys.\ J.\ C {\bf 78}, 938 (2018).
  
  
  \bibitem{water}
  Rana~A.~Fine, Frank~J.~Millero,
  J.\ Chem.\ Phys. {\bf 59}, 5529 (1973).
 
  \bibitem{mercury}
  Hugh D. Young; Roger A. Freedman,  University Physics with Modern Physics, Addison-Wesley, pp. 356 (2012).
 
  
\end{thebibliography}
 \end{document}